\documentclass[aps,pra,showpacs,twocolumn]{revtex4-1}
\usepackage{amsmath}
\usepackage{amssymb}
\usepackage{graphicx}
\usepackage{epstopdf}
\usepackage{times,txfonts}
\usepackage{easybmat}
\usepackage{mathtools}
\newcommand{\ket}[1]{|#1\rangle}
\newcommand{\bra}[1]{\langle#1|}
\usepackage[bookmarks=false]{hyperref}
\hypersetup{colorlinks=true,citecolor=blue,linkcolor=blue,urlcolor=blue,pdfstartview=FitH,bookmarksopen=true}
\usepackage{color,soul}

\begin{document}

\title{Dynamical-decoupling-protected nonadiabatic holonomic quantum computation}
\author{P. Z. Zhao}
\affiliation{Department of Physics, Shandong University, Jinan 250100, China}
\author{X. Wu}
\affiliation{Department of Physics, Shandong University, Jinan 250100, China}
\author{D. M. Tong}
\email{tdm@sdu.edu.cn}
\affiliation{Department of Physics, Shandong University, Jinan 250100, China}
\date{\today}

\begin{abstract}
The main obstacles to the realization of high-fidelity quantum gates are the control errors arising from inaccurate manipulation of a quantum system and the decoherence caused by the interaction between the quantum system and its environment. Nonadiabatic holonomic quantum computation allows for high-speed implementation of whole-geometric quantum gates, making quantum computation robust against control errors. Dynamical decoupling provides an effective method to protect quantum gates against environment-induced decoherence, regardless of collective decoherence or independent decoherence. In this paper, we put forward a protocol of nonadiabatic holonomic quantum computation protected by dynamical decoupling . Due to the combination of nonadiabatic holonomic quantum computation and dynamical decoupling, our protocol not only possesses the intrinsic robustness against control errors but also protects quantum gates against environment-induced decoherence.
\end{abstract}
\maketitle

\section{Introduction}

Quantum computation provides an effective solution to certain problems, such as factoring large integers \cite{Shor} and searching unsorted data \cite{Grover}. The implementation of circuit-based quantum computation relies on the ability to realize a universal set of high-fidelity quantum gates, including arbitrary one-qubit gates and a nontrivial two-qubit gate \cite{Bremner}. However, there are two main obstacles to the realization of high-fidelity quantum gates. One is the control errors arising from inaccurate manipulation of quantum systems. The other one is the decoherence caused by the interaction between the quantum system and its environment. Geometric phases are only dependent on  evolution paths of quantum systems but independent of the evolution details, and therefore quantum computation based on geometric phases is robust against control errors.

In 1984, Berry found that a quantum system in a nondegenerate eigenstate undergoing adiabatic and cyclic evolution can acquire a geometric phase in addition to a dynamical phase \cite{Berry}. The notion of geometric phases was then extended to quantum systems in degenerate eigenstates \cite{Wilczek}, in nonadiabatic evolution \cite{Aharonov,Anandan}, and in mixed states \cite{Uhlmann,Sjoqvist2000,Tong2004}. Until now, pure-state geometric phases have been used to realize quantum computation while mixed-state geometric phases have not.
The early proposals \cite{Zanardi,Jones,Duan} of geometric quantum computation are based on adiabatic Abelian geometric phases \cite{Berry} and adiabatic non-Abelian geometric phases \cite{Wilczek}. However, these proposals require quantum systems to undergo adiabatic evolution, which makes quantum systems evolve for a long time. To resolve this problem, nonadiabatic geometric quantum computation \cite{Wang,Zhu} based on nonadiabatic Abelian geometric phases \cite{Aharonov} and nonadiabatic holonomic quantum computation \cite{Sjoqvist,Xu} based on nonadiabatic non-Abelian geometric phases \cite{Anandan} were proposed.
Compared with nonadiabatic geometric quantum computation that uses the geometric phase as one parameter of a quantum gate, nonadiabatic holonomic quantum computation uses the holonomic matrix itself as a quantum gate. This makes nonadiabatic holonomic quantum computation possess a whole-geometric property. Due to the merits of both geometric robustness and high-speed implementation without the limit of adiabatic evolution, nonadiabatic holonomic quantum computation has received increasing attention.

The first protocol of nonadiabatic holonomic quantum computation is based on a three-level quantum system driven by two resonant laser pulses \cite{Sjoqvist}. It needs to combine two one-qubit gates to realize an arbitrary one-qubit gate. To simplify the operations, the single-shot protocol of nonadiabatic holonomic quantum computation \cite{Xu2015,Sjoqvist2016} and the single-loop protocol of nonadiabatic holonomic quantum computation \cite{Herterich} were proposed. The improved protocols allow us to realize an arbitrary one-qubit gate by a single-shot implementation and thus reduce the exposure time of nonadiabatic holonomic gates to error sources. To further shorten the exposure time of quantum gates to error sources, the path-shortening protocol of nonadiabatic holonomic quantum computation was put forward \cite{Xu2018}, where nonadiabatic holonomic gates can be realized based on a class of extended evolution paths that are shorter than the former ones.
The key to realizing nonadiabatic holonomic quantum computation based on these protocols is to find the Hamiltonians that make the quantum system satisfy both the cyclic evolution condition and the parallel transport condition.
Recently, a general approach of constructing Hamiltonians for nonadiabatic holonomic quantum computation was put forward \cite{Zhao}. By using this approach, one can easily find a Hamiltonian making the quantum system evolve along a desired path so that nonadiabatic holonomic gates can be realized with an economical evolution time. Up to now, a lot of works both in theories \cite{Johansson,Spiegelberg,Mousolou,Sjoqvist2015,Wang2016,Xu2017,XuGF,Mousolou2017,Mousolou2018,Zhao2017,Zhao2018,Xia,
Chen,Zhao2019,Ramberg,Xing} and experiments \cite{Feng,Li,Abdumalikov,Sun,Danilin,Zhang2019,Egger,Zu,Camejo,
Sekiguchi,Zhou,Nagata,Isida,Long} have contributed to nonadiabatic holonomic quantum computation.

While some protocols tried to reduce the influence of decoherence by shorting the exposure time of quantum gates, another line of protecting quantum gates against decoherence is to use decoherence-mitigation methods.
To make quantum gates robust against both control errors and decoherence, the combination of nonadiabatic holonomic quantum computation and decoherence-mitigation methods is a promising strategy. The first protocol of nonadiabatic holonomic quantum computation in decoherence-free subspaces was put forward in Ref. \cite{Xu}. Afterwards, a number of alternative protocols \cite{Liang,Zhang,ZhaoPZ,Wang2018} and physical implementation schemes \cite{Zhou2015,Xue,Xue2016,Liu2017,Zhu2019} were put forward. These proposals are based on the system-environment interaction being with some symmetry and they focus mainly on protecting nonadiabatic holonomic gates against collective decoherence, especially the collective dephasing.

In this paper, we propose a protocol of nonadiabatic holonomic quantum computation protected by dynamical decoupling. The decoherence-mitigation method used here is dynamical decoupling \cite{Viola}, and therefore there is no need to require system-environment interaction to have some symmetry. Due to the combination of nonadiabatic holonomic quantum computation and dynamical decoupling, our protocol not only possesses the intrinsic robustness against control errors but also protects quantum gates against environment-induced decoherence, regardless of collective decoherence or independent decoherence.

\section{Physical model}

Before proceeding further, we briefly review the basic idea of nonadiabatic holonomic quantum computation \cite{Sjoqvist,Xu}. Consider an $N$-dimensional quantum system exposed to the Hamiltonian $H(t)$.  Assume there is
an $L-$dimensional subspace $\mathcal{S}(t)=\mathrm{Span}\{\ket{\phi_{k}(t)}\}^{L}_{k=1}$, where $\ket{\phi_{k}(t)}$ are orthonormal basis vectors and satisfy the Schr\"{o}dinger equation $i\ket{\dot{\phi}_{k}(t)}=H(t)\ket{\phi_{k}(t)}$.
If $\ket{\phi_{k}(t)}$ satisfy the following conditions
\begin{align}\label{eq}
&(\mathrm{i})~ \sum^{L}_{k=1}\ket{\phi_{k}(\tau)}\bra{\phi_{k}(\tau)}=
\sum^{L}_{k=1}\ket{\phi_{k}(0)}\bra{\phi_{k}(0)}, \notag\\
&(\mathrm{ii})~~\bra{\phi_{k}(t)}H(t)\ket{\phi_{l}(t)}=0,~~~~k,l=1,\cdot\cdot\cdot L,
\end{align}
then the unitary operator $U(\tau)$ with $\ket{\phi_{k}(\tau)}=U(\tau)\ket{\phi_{k}(0)}$ is a nonadiabatic holonomic gate acting on $\mathcal{S}(0)$. Here, $\tau$ is the evolution period.

Let us now elucidate our physical model. We consider a quantum system consisting of $N$ physical qubits, which interact though the well-known $XXZ$ coupling \cite{Yang,Johnson,Duan2003,Alcaraz,Canosa,Breunig}. The Hamiltonian reads
\begin{align}\label{eq1}
H=\sum_{k<l}\left[J^{x}_{kl}\left(\sigma^{x}_{k}\sigma^{x}_{l}+\sigma^{y}_{k}\sigma^{y}_{l}\right)
+J^{z}_{kl}\sigma^{z}_{k}\sigma^{z}_{l}\right],
\end{align}
where $J^{x}_{kl}$ and $J^{z}_{kl}$ are the real-valued controllable coupling parameters and $\sigma^{\alpha}_{m}$ represent  the Pauli $\alpha$ operators ($\alpha=x,y,z$) acting on the $m$th qubit ($m=k,l$). For the quantum system considered here, we assume that each physical qubit interacts independently with its environment. The interaction Hamiltonian reads
\begin{align}\label{eq2}
H_{I}=\sum_{k,\alpha}\sigma^{\alpha}_{k}\otimes B^{\alpha}_{k},
\end{align}
where $B^{\alpha}_{k}$ is the environment operator corresponding to the system operator $\sigma^{\alpha}_{k}$.
If $B^{\alpha}_{k}$ is independent of the qubit index $k$, the environment-induced decoherence is reduced to collective decoherence. In particular, if the system operator is further taken as $\sigma^{z}_{k}$, then the collective decoherence yields collective dephasing. To protect nonadiabatic holonomic gates against collective dephasing,
nonadiabatic holonomic quantum computation in decoherence-free subspaces was proposed \cite{Xu}, and to protect nonadiabatic holonomic gates against collective decoherence, nonadiabatic holonomic quantum computation in noiseless systems was proposed \cite{Zhang}. For the more complicated decoherence induced by the interaction $H_{I}$ in Eq. (\ref{eq2}), dynamical decoupling  provides an effective method to protect nonadiabatic holonomic gates against decoherence.

Dynamical decoupling operates by applying a periodic sequence of fast and strong symmetrizing pulses to quantum
systems to suppress the effect of undesired system-environment interaction. For the system-environment interaction in Eq. (\ref{eq2}), we can use a periodic sequence with the decoupling operations $\{\otimes^{N}_{k=1}I_{k},\otimes^{N}_{k=1}\sigma^{x}_{k},\otimes^{N}_{k=1}\sigma^{y}_{k},
\otimes^{N}_{k=1}\sigma^{z}_{k}\}$ to suppress its effect. The corresponding unitary operator over a period of time reads
\begin{widetext}
\begin{align}\label{req}
U_{I}=&\left[\left(\otimes^{N}_{k=1}\sigma^{z}_{k}\right)e^{-i H_{I}\tau} \left(\otimes^{N}_{k=1}\sigma^{z}_{k}\right)\right]
\left[\left(\otimes^{N}_{k=1}\sigma^{y}_{k}\right)e^{-i H_{I}\tau} \left(\otimes^{N}_{k=1}\sigma^{y}_{k}\right)\right]
\left[\left(\otimes^{N}_{k=1}\sigma^{x}_{k}\right)e^{-i H_{I}\tau} \left(\otimes^{N}_{k=1}\sigma^{x}_{k}\right)\right]
\left[\left(\otimes^{N}_{k=1}I_{k}\right)e^{-i H_{I}\tau} \left(\otimes^{N}_{k=1}I_{k}\right)\right]
\notag\\
=&e^{-i\left(\otimes^{N}_{k=1}\sigma^{z}_{k}\right)H_{I}\left(\otimes^{N}_{k=1}\sigma^{z}_{k}\right)\tau}
e^{-i\left(\otimes^{N}_{k=1}\sigma^{y}_{k}\right)H_{I}\left(\otimes^{N}_{k=1}\sigma^{y}_{k}\right)\tau}
e^{-i\left(\otimes^{N}_{k=1}\sigma^{x}_{k}\right)H_{I}\left(\otimes^{N}_{k=1}\sigma^{x}_{k}\right)\tau}
e^{-iH_{I}\tau}
\notag\\
=&e^{-i\left[\left(\otimes^{N}_{k=1}\sigma^{z}_{k}\right)H_{I}\left(\otimes^{N}_{k=1}\sigma^{z}_{k}\right)
+\left(\otimes^{N}_{k=1}\sigma^{y}_{k}\right)H_{I}\left(\otimes^{N}_{k=1}\sigma^{y}_{k}\right)
+\left(\otimes^{N}_{k=1}\sigma^{x}_{k}\right)H_{I}\left(\otimes^{N}_{k=1}\sigma^{x}_{k}\right)
+H_{I}\right]\tau}+O(\tau^2)
\notag\\
=&e^{-i\sum^{N}_{k=1}(-\sigma_{k}^{x}\otimes B_{k}^{x}-\sigma_{k}^{y}\otimes B_{k}^{y}+\sigma_{k}^{z}\otimes B_{k}^{z}
-\sigma_{k}^{x}\otimes B_{k}^{x}+\sigma_{k}^{y}\otimes B_{k}^{y}-\sigma_{k}^{z}\otimes B_{k}^{z}
+\sigma_{k}^{x}\otimes B_{k}^{x}-\sigma_{k}^{y}\otimes B_{k}^{y}-\sigma_{k}^{z}\otimes B_{k}^{z}
+\sigma_{k}^{x}\otimes B_{k}^{x}+\sigma_{k}^{y}\otimes B_{k}^{y}+\sigma_{k}^{z}\otimes B_{k}^{z})\tau}
+O(\tau^2)
\notag\\
=&\otimes^{N}_{k=1}I_{k}+O(\tau^2),
\end{align}
\end{widetext}
where $\tau$ is the duration time of pulse intervals and $I_{k}$ is the identity operator acting on the $k$th qubit.
This result indicates that up to the first-order term $O(\tau)$, the system-environment interaction can be completely eliminated by using a decoupling pulse sequence.

To realize dynamical-decoupling-protected nonadiabatic holonomic quantum computation, the decoupling pulse sequence needs to be inserted into the native dynamical evolution of the quantum system.
Therefore, we need to properly choose the Hamiltonian that not only makes the decoupling pulse sequence compatible with the dynamical evolution but also keeps the cyclic evolution condition as well as the parallel transport condition valid.
To this end, we choose the Hamiltonian in Eq. (\ref{eq1}), which commutes with the decoupling operations. In this case, we can realize the desired evolution protected by dynamical decoupling.

\section{Implementation}

To perform dynamical-decoupling-protected nonadiabatic holonomic quantum computation, we need to realize a universal set of quantum gates, including arbitrary one-qubit gates and a nontrivial two-qubit gate.

First, we realize an arbitrary one-qubit gate. To complete our realization, we utilize three physical qubits to encode a logical qubit. The specific encoding is
\begin{align}\label{req}
\ket{0}_{L}=\ket{001},~~\ket{1}_{L}=\ket{010}.
\end{align}
Meanwhile, we use $\ket{a}=\ket{100}$ as an auxiliary state.
In this case, we can apply the periodic sequence with decoupling operations  $\{\otimes^{3}_{k=1}I_{k},\otimes^{3}_{k=1}\sigma^{x}_{k},
\otimes^{3}_{k=1}\sigma^{y}_{k},\otimes^{3}_{k=1}\sigma^{z}_{k}\}$ to the quantum system to protect quantum information against decoherence.

To realize nonadiabatic holonomic gates, we set the nonzero parameters of the Hamiltonian in Eq. (\ref{eq1}) to be
\begin{align}
&J^{x}_{12}=-\frac{J_{1}(t)}{2}\cos\phi_{1}\cos\frac{\theta_{1}}{2},~~
J^{x}_{13}=\frac{J_{1}(t)}{2}\cos\phi_{1}\sin\frac{\theta_{1}}{2},
\notag\\
&J^{z}_{23}=J_{1}(t)\sin\phi_{1},
\end{align}
where $J_{1}(t)$ is a time-dependent parameter, and $\phi_{1}$ and $\theta_{1}$ are time-independent parameters.
In this case, the Hamiltonian reads
\begin{align}\label{req1}
H_{1}(t)=&\frac{J_{1}(t)}{2}\cos\phi_{1}\bigg[-\cos\frac{\theta_{1}}{2}
\left(\sigma^{x}_{1}\sigma^{x}_{2}+\sigma^{y}_{1}\sigma^{y}_{2}\right)
\notag\\
&+\sin\frac{\theta_{1}}{2}\left(\sigma^{x}_{1}\sigma^{x}_{3}+\sigma^{y}_{1}\sigma^{y}_{3}\right)\bigg]
+J_{1}(t)\sin\phi_{1}\sigma^{z}_{2}\sigma^{z}_{3}.
\end{align}
By using the basis $\{\ket{0}_{L},\ket{1}_{L},\ket{a}\}$, this Hamiltonian can be recast as
\begin{align}
H_{1}(t)=&J_{1}(t)\cos\phi_{1}\left(\sin\frac{\theta_{1}}{2}\ket{a}_{L}\bra{0}-\cos\frac{\theta_{1}}{2}\ket{a}_{L}\bra{1}
+\mathrm{H.c.}\right)
\notag\\
&+J_{1}(t)\sin\phi_{1}(\ket{a}\bra{a}-\ket{0}_{LL}\bra{0}-\ket{1}_{LL}\bra{1}),
\end{align}
which can be further rewritten as
\begin{align}\label{eq3}
H_{1}(t)=&J_{1}(t)\cos\phi_{1}\left(\sin\frac{\theta_{1}}{2}\ket{a}_{L}\bra{0}-\cos\frac{\theta_{1}}{2}\ket{a}_{L}\bra{1}
+\mathrm{H.c.}\right)
\notag\\
&+2J_{1}(t)\sin\phi_{1}\ket{a}\bra{a}
\notag\\
&-J_{1}(t)\sin\phi_{1}(\ket{a}\bra{a}+\ket{0}_{LL}\bra{0}+\ket{1}_{LL}\bra{1}).
\end{align}
It is noteworthy that $\ket{a}\bra{a}+\ket{0}_{LL}\bra{0}+\ket{1}_{LL}\bra{1}$ is an identity operator and thus $-J_{1}(t)\sin\phi_{1}(\ket{a}\bra{a}+\ket{0}_{LL}\bra{0}+\ket{1}_{LL}\bra{1})$ can only generate a global phase
during evolution. This global phase does not affect the quantum gates and therefore the terms $-J_{1}(t)\sin\phi_{1}(\ket{a}\bra{a}+\ket{0}_{LL}\bra{0}+\ket{1}_{LL}\bra{1})$ in Eq. (\ref{eq3}) can be ignored. If we introduce two orthonormal states,
\begin{align}
\ket{d}&=\cos\frac{\theta_{1}}{2}\ket{0}_L+\sin\frac{\theta_{1}}{2}\ket{1}_L,
\notag\\
\ket{b}&=\sin\frac{\theta_{1}}{2}\ket{0}_L-\cos\frac{\theta_{1}}{2}\ket{1}_L,
\end{align}
the Hamiltonian then reads
\begin{align}
H_{1}(t)=&J_{1}(t)\cos\phi_{1}(\ket{a}\bra{b}+\ket{b}\bra{a})
+2J_{1}(t)\sin\phi_{1}\ket{a}\bra{a}.
\end{align}
The evolution operator corresponding to the above Hamiltonian can be written as $U_{1}(t)=\exp[-i\int^{t}_{0}H_{1}(t^{\prime})dt^{\prime}]$, which can be explicitly expressed as
\begin{align}\label{eq4}
U_{1}(t)=&\ket{d}\bra{d}+e^{-i\int^{t}_{0}J_{1}(t^{\prime})dt^{\prime}\sin\phi_{1}(\ket{a}\bra{a}+\ket{b}\bra{b})}
\notag\\
&\times e^{-i\int^{t}_{0}J_{1}(t^{\prime})dt^{\prime}[\cos\phi_{1}(\ket{a}\bra{b}+\ket{b}\bra{a})+
\sin\phi_{1}(\ket{a}\bra{a}-\ket{b}\bra{b})]}.
\end{align}
If the evolution time $T$ is taken to satisfy
\begin{align}
\int^{T}_{0}J_{1}(t)dt=\pi,
\end{align}
then the evolution operator is reduced to
\begin{align}\label{eq5}
U_{1}(T)=\ket{d}\bra{d}+e^{-i(\pi+\pi\sin\phi_{1})}\ket{b}\bra{b}+e^{-i(\pi+\pi\sin\phi_{1})}\ket{a}\bra{a}.
\end{align}
From Eqs. (\ref{eq4}) and (\ref{eq5}), we can see that a state initially prepared in the computational space $\mathcal{S}_{1}=\mathrm{Span}\{\ket{0}_{L},\ket{1}_{L}\}$ will evolve outside $\mathcal{S}_{1}$ during $t\in(0,T)$ and then return back to $\mathcal{S}_{1}$ at $t=T$. Thus, the cyclic evolution condition (i) is satisfied. By using the commutation relation $[H_{1}(t),U_{1}(t)]=0$, we can verify that $\bra{\phi(t)}H_{1}(t)\ket{\phi(t)}
=\bra{\phi(0)}U^{\dagger}_{1}(t)H_{1}(t)U_{1}(t)\ket{\phi(0)}
=\bra{\phi(0)}H_{1}(t)\ket{\phi(0)}=0$, where $\ket{\phi(t)}$ is an evolution state such that $\ket{\phi(t)}=U_{1}(t)\ket{\phi(0)}$ with $\ket{\phi(0)}\in\mathcal{S}_{1}$. It indicates that the parallel transport condition (ii) is satisfied. Therefore, $U_{1}(T)$ is a holonomic transformation. Acting on the computational space $\mathcal{S}_{1}$, the evolution operator $U_{1}(T)$
is equivalent to
\begin{align}
U_{1}=\ket{d}\bra{d}+e^{-i(\pi+\pi\sin\phi_{1})}\ket{b}\bra{b},
\end{align}
which plays the role of a nonadiabatic holonomic gate.

In the following, we demonstrate how to use dynamical decoupling to protect the dynamical evolution for the realization of nonadiabatic holonomic gates. We assume that the quantum system is coupled to its environment with the total Hamiltonian $\mathcal{H}(t)=H_{1}(t)+H_{E}+H_{I}$, where $H_{1}(t)$ is the system Hamiltonian in Eq. (\ref{req1}),
$H_{E}$ is the environment Hamiltonian, and $H_{I}$ is the interaction Hamiltonian in Eq. (\ref{eq2}). If the decoupling operations  $\{\otimes^{3}_{k=1}I_{k},\otimes^{3}_{k=1}\sigma^{x}_{k},
\otimes^{3}_{k=1}\sigma^{y}_{k},\otimes^{3}_{k=1}\sigma^{z}_{k}\}$ are applied to the quantum system, the unitary operator reads
\begin{align}
\mathcal{U}(4\tau)=&\left[\left(\otimes^{3}_{k=1}\sigma^{z}_{k}\right)e^{-i\int^{4\tau}_{3\tau}\mathcal{H}(t)dt} \left(\otimes^{3}_{k=1}\sigma^{z}_{k}\right)\right]
\notag\\
&\times
\left[\left(\otimes^{3}_{k=1}\sigma^{y}_{k}\right)e^{-i\int^{3\tau}_{2\tau}\mathcal{H}(t)dt} \left(\otimes^{3}_{k=1}\sigma^{y}_{k}\right)\right]
\notag\\
&\times
\left[\left(\otimes^{3}_{k=1}\sigma^{x}_{k}\right)e^{-i\int^{2\tau}_{\tau}\mathcal{H}(t)dt} \left(\otimes^{3}_{k=1}\sigma^{x}_{k}\right)\right]
\notag\\
&\times
\left[\left(\otimes^{3}_{k=1}I_{k}\right)e^{-i\int^{\tau}_{0}\mathcal{H}(t)dt} \left(\otimes^{3}_{k=1}I_{k}\right)\right]
\notag\\
=&e^{-i\left(\otimes^{3}_{k=1}\sigma^{z}_{k}\right)\int^{4\tau}_{3\tau}\mathcal{H}(t)dt
\left(\otimes^{3}_{k=1}\sigma^{z}_{k}\right)}
e^{-i\left(\otimes^{3}_{k=1}\sigma^{y}_{k}\right)\int^{3\tau}_{2\tau}\mathcal{H}(t)dt
\left(\otimes^{3}_{k=1}\sigma^{y}_{k}\right)}
\notag\\
&\times
e^{-i\left(\otimes^{3}_{k=1}\sigma^{x}_{k}\right)\int^{2\tau}_{\tau}\mathcal{H}(t)dt
\left(\otimes^{3}_{k=1}\sigma^{x}_{k}\right)}
e^{-i\int^{\tau}_{0}\mathcal{H}(t)dt}
\notag\\
=&e^{-i\left[\left(\otimes^{3}_{k=1}\sigma^{z}_{k}\right)\int^{4\tau}_{3\tau}H_{1}(t)dt\left(\otimes^{3}_{k=1}\sigma^{z}_{k}\right)
+\left(\otimes^{3}_{k=1}\sigma^{z}_{k}\right)H_{I}\left(\otimes^{3}_{k=1}\sigma^{z}_{k}\right)\tau+H_{E}\tau\right]}
\notag\\
&\times
e^{-i\left[\left(\otimes^{3}_{k=1}\sigma^{y}_{k}\right)\int^{3\tau}_{2\tau}H_{1}(t)dt\left(\otimes^{3}_{k=1}\sigma^{y}_{k}\right)
+\left(\otimes^{3}_{k=1}\sigma^{y}_{k}\right)H_{I}\left(\otimes^{3}_{k=1}\sigma^{y}_{k}\right)\tau+H_{E}\tau\right]}
\notag\\
&\times
e^{-i\left[\left(\otimes^{3}_{k=1}\sigma^{x}_{k}\right)\int^{2\tau}_{\tau}H_{1}(t)dt\left(\otimes^{3}_{k=1}\sigma^{x}_{k}\right)
+\left(\otimes^{3}_{k=1}\sigma^{x}_{k}\right)H_{I}\left(\otimes^{3}_{k=1}\sigma^{x}_{k}\right)\tau+H_{E}\tau\right]}
\notag\\
&\times
e^{-i\left[\int^{\tau}_{0}H_{1}(t)dt
+H_{I}\tau+H_{E}\tau\right]}.
\end{align}
Since the chosen Hamiltonian $H_{1}(t)$ commutes with the decoupling operations, the unitary operator can be recast as
\begin{align}
\mathcal{U}(4\tau)=&e^{-i\left[\int^{4\tau}_{0}H_{1}(t)dt
+\sum_{\alpha=x,y,z}\left(\otimes^{3}_{k=1}\sigma^{\alpha}_{k}\right)H_{I}\left(\otimes^{3}_{k=1}\sigma^{\alpha}_{k}\right)\tau
+H_{I}\tau+4H_{E}\tau\right]}+O(\tau^2)
\notag\\
=&e^{-i\big[\int^{4\tau}_{0}H_{1}(t)dt+4H_{E}\tau\big]}+O(\tau^2)
\notag\\
=&e^{-i\int^{4\tau}_{0}H_{1}(t)dt}\otimes e^{-i4H_{E}\tau}+O(\tau^2)
\notag\\
=&U_{1}(4\tau)\otimes U_{E}(4\tau)+O(\tau^2),
\end{align}
where $U_{1}(4\tau)$ and $U_{E}(4\tau)$ are the evolution operators of the quantum system and its environment, and the relation $\sum_{\alpha=x,y,z}(\otimes^{3}_{k=1}\sigma^{\alpha}_{k})H_{I}(\otimes^{3}_{k=1}\sigma^{\alpha}_{k})
\tau+H_{I}\tau=0$ has been utilized to derive the second line. From the above result, one can see that up to the first-order term $O(\tau)$, the quantum system is completely decoupled from its environment at $t=4\tau$.
If we choose $\tau\ll T$, then we can repeat the above process over and over so that the effect of system-environment interaction will be suppressed in each interval of duration $4\tau$.
Therefore, we can use dynamical decoupling to protect nonadiabatic holonomic gates.
On the contrary, if we do not use dynamical decoupling to protect quantum gates, the unitary operator will be expressed as
\begin{align}\label{req2}
\mathcal{U}(4\tau)=&e^{-i\int^{4\tau}_{0}\mathcal{H}(t)dt}
\notag\\
=&e^{-i\left[\int^{4\tau}_{0}H_{1}(t)dt+4H_{I}\tau+4H_{E}\tau\right]}
\notag\\
=&\Big[e^{-i\int^{4\tau}_{0}H_{1}(t)dt}\otimes I_{E}\Big]
\left(e^{-i4H_{I}\tau}\right)
\left(I_{S}\otimes e^{-i4H_{E}\tau}\right)+O(\tau^2)
\notag\\
=&[U_{1}(4\tau)\otimes I_{E}]U_{I}(4\tau)[I_{S}\otimes U_{E}(\tau)]+O(\tau^2),
\end{align}
where $U_{I}(4\tau)$ is the evolution operator induced by the interaction Hamiltonian $H_{I}$, and $I_{S(E)}$ is the identity operator of the quantum system (environment).
From Eqs. (\ref{eq2}) and (\ref{req2}), we can obviously conclude that the system-environment interaction will affect the native dynamical evolution of the quantum system.

Finally, we demonstrate that an arbitrary one-qubit gate can be realized by using the nonadiabatic holonomic gate $U_{1}$. One can see that $U_{1}$ can be rewritten as
\begin{align}
U_{1}=e^{i\gamma_{1}/2}e^{-i\gamma_{1}(\sin\theta_{1}X+\cos\theta_{1}Z)/2},
\end{align}
where $\gamma_{1}=-(\pi+\pi\sin\phi_{1})$, and $X$ and $Z$ are the Pauli $x$ operator and Pauli $z$ operator acting on $\ket{0}_{L}$ and $\ket{1}_{L}$, respectively. Ignoring a trivial global phase, we can obviously see that $U_{1}$ is a quantum gate with a rotation axis in the $x-z$ plane and the rotation angle $\gamma_{1}$.
An arbitrary one-qubit gate can be realized by combining two such quantum gates about unparallel axes in the plane. For
example, $U_{1}$ is reduced to the quantum gate about the $x$ axis by setting $\theta_{1}=\pi/2$ and about the $z$ axis by setting $\theta_{1}=0$. By combining these noncommuting one-qubit gates, an arbitrary one-qubit gate can be realized.

Second, we realize a nontrivial two-qubit gate. To this end, we use six physical qubits to encode two logical qubits. To make the two-logical-qubit encoding compatible with the one-logical-qubit encoding, we encode two-logical-qubit states as
\begin{align}
\ket{00}_{L}&=\ket{001001},~\ket{01}_{L}=\ket{001010},
\notag\\
\ket{10}_{L}&=\ket{010001},~\ket{11}_{L}=\ket{010010}.
\end{align}
Meanwhile, we use $\ket{a_{1}}=\ket{011000}$ and $\ket{a_{2}}=\ket{000011}$ as auxiliary states.
In this case, we can utilize decoupling operations $\{\otimes^{6}_{k=1}I_{k},\otimes^{6}_{k=1}\sigma^{x}_{k},
\otimes^{6}_{k=1}\sigma^{y}_{k},\otimes^{6}_{k=1}\sigma^{z}_{k}\}$ to suppress the effect of the undesired system-environment interaction in Eq. (\ref{eq2}).

To realize nonadiabatic holonomic gates, we take the nonzero parameters of the Hamiltonian in Eq. (\ref{eq1}) as
\begin{align}
&J^{x}_{25}=-\frac{J_{2}(t)}{2}\cos\phi_{2}\cos\frac{\theta_{2}}{2},~~
J^{x}_{26}=\frac{J_{2}(t)}{2}\cos\phi_{2}\sin\frac{\theta_{2}}{2},
\notag\\
&J^{z}_{23}=J_{2}(t)\sin\phi_{2},
\end{align}
where $J_{2}(t)$ is time dependent, and $\phi_{2}$ and $\theta_{2}$ are time independent.
In this case, the Hamiltonian can be expressed as
\begin{align}
H_{2}(t)=&\frac{J_{2}(t)}{2}\cos\phi_{2}\bigg[-\cos\frac{\theta_{2}}{2}
\left(\sigma^{x}_{2}\sigma^{x}_{5}+\sigma^{y}_{2}\sigma^{y}_{5}\right)
\notag\\
&+\sin\frac{\theta_{2}}{2}\left(\sigma^{x}_{2}\sigma^{x}_{6}+\sigma^{y}_{2}\sigma^{y}_{6}\right)\bigg]
+J_{2}(t)\sin\phi_{2}\sigma^{z}_{2}\sigma^{z}_{3}.
\end{align}
By using the basis $\{\ket{00}_{L},\ket{01}_{L},\ket{10}_{L},\ket{11}_{L},\ket{a_{1}},\ket{a_{2}}\}$,
this Hamiltonian can be recast as
\begin{align}
H_{2}(t)=&J_{2}(t)\cos\phi_{2}\bigg[\left(\sin\frac{\theta_{2}}{2}\ket{a_{1}}_{L}\bra{00}
-\cos\frac{\theta_{2}}{2}\ket{a_{1}}_{L}\bra{01}+\mathrm{H.c.}\right)
\notag\\
&-\left(\cos\frac{\theta_{2}}{2}\ket{a_{2}}_{L}\bra{10}
-\sin\frac{\theta_{2}}{2}\ket{a_{2}}_{L}\bra{11}+\mathrm{H.c.}\right)\bigg]
\notag\\
&+J_{2}(t)\sin\phi_{2}[(\ket{a_{1}}\bra{a_{1}}-\ket{00}_{LL}\bra{00}-\ket{01}_{LL}\bra{01})
\notag\\
&+(\ket{a_{2}}\bra{a_{2}}-\ket{10}_{LL}\bra{10}-\ket{11}_{LL}\bra{11})],
\end{align}
which can be further recast as
\begin{align}\label{eq9}
H_{2}(t)=&J_{2}(t)\cos\phi_{2}\bigg[\left(\sin\frac{\theta_{2}}{2}\ket{a_{1}}_{L}\bra{00}
-\cos\frac{\theta_{2}}{2}\ket{a_{1}}_{L}\bra{01}+\mathrm{H.c.}\right)
\notag\\
&-\left(\cos\frac{\theta_{2}}{2}\ket{a_{2}}_{L}\bra{10}
-\sin\frac{\theta_{2}}{2}\ket{a_{2}}_{L}\bra{11}+\mathrm{H.c.}\right)\bigg]
\notag\\
&+2J_{2}(t)\sin\phi_{2}(\ket{a_{1}}\bra{a_{1}}+\ket{a_{2}}\bra{a_{2}})
\notag\\
&-J_{2}(t)\sin\phi_{2}(\ket{a_{1}}\bra{a_{1}}+\ket{a_{2}}\bra{a_{2}}+\ket{00}_{LL}\bra{00}
\notag\\
&+\ket{01}_{LL}\bra{01}+\ket{10}_{LL}\bra{10}+\ket{11}_{LL}\bra{11}).
\end{align}
It is noteworthy that $\ket{a_{1}}\bra{a_{1}}+\ket{a_{2}}\bra{a_{2}}+\ket{00}_{LL}\bra{00}
+\ket{01}_{LL}\bra{01}+\ket{10}_{LL}\bra{10}+\ket{11}_{LL}\bra{11}$ is an identity operator and thus $-J_{2}(t)\sin\phi_{2}(\ket{a_{1}}\bra{a_{1}}+\ket{a_{2}}\bra{a_{2}}+\ket{00}_{LL}\bra{00}
+\ket{01}_{LL}\bra{01}+\ket{10}_{LL}\bra{10}+\ket{11}_{LL}\bra{11})$ can only generate a global phase
during evolution. This global phase does not affect the quantum gates and therefore the terms $-J_{2}(t)\sin\phi_{2}(\ket{a_{1}}\bra{a_{1}}+\ket{a_{2}}\bra{a_{2}}+\ket{00}_{LL}\bra{00}
+\ket{01}_{LL}\bra{01}+\ket{10}_{LL}\bra{10}+\ket{11}_{LL}\bra{11})$ in Eq. (\ref{eq9}) can be ignored.
If we introduce four orthonormal states,
\begin{align}
\ket{d_{1}}&=\cos\frac{\theta_{2}}{2}\ket{00}_{L}+\sin\frac{\theta_{2}}{2}\ket{01}_{L},
\notag\\
\ket{b_{1}}&=\sin\frac{\theta_{2}}{2}\ket{00}_{L}-\cos\frac{\theta_{2}}{2}\ket{01}_{L},
\notag\\
\ket{d_{2}}&=\sin\frac{\theta_{2}}{2}\ket{10}_{L}+\cos\frac{\theta_{2}}{2}\ket{11}_{L},
\notag\\
\ket{b_{2}}&=\cos\frac{\theta_{2}}{2}\ket{10}_{L}-\sin\frac{\theta_{2}}{2}\ket{11}_{L},
\end{align}
the Hamiltonian can be further written as
\begin{align}
H_{2}(t)=&J_{2}(t)\cos\phi_{2}\left(\ket{a_{1}}\bra{b_{1}}+\ket{b_{1}}\bra{a_{1}}\right)
+2J_{2}(t)\sin_{2}\phi_{2}\ket{a_{1}}\bra{a_{1}}
\notag\\
&-J_{2}(t)\cos\phi_{2}\left(\ket{a_{2}}\bra{b_{2}}+\ket{b_{2}}\bra{a_{2}}\right)
\notag\\
&+2J_{2}(t)\sin\phi_{2}\ket{a_{2}}\bra{a_{2}}.
\end{align}
The evolution operator corresponding to this Hamiltonian then reads $U_{2}(t)=\exp[-i\int^{t}_{0}H_{2}(t^{\prime})dt^{\prime}]$, which can be explicitly expressed as
\begin{align}\label{eq7}
U_{2}(t)=&\ket{d_{1}}\bra{d_{1}}
+e^{-i\int^{t}_{0}J_{2}(t^{\prime})dt^{\prime}\sin\phi_{2}(\ket{a_{1}}\bra{a_{1}}+\ket{b_{1}}\bra{b_{1}})}
\notag\\
&\times e^{-i\int^{t}_{0}J_{2}(t^{\prime})dt^{\prime}[\cos\phi_{2}(\ket{a_{1}}\bra{b_{1}}+\ket{b_{1}}\bra{a_{1}})+
\sin\phi_{2}(\ket{a_{1}}\bra{a_{1}}-\ket{b_{1}}\bra{b_{1}})]}
\notag\\
&+\ket{d_{2}}\bra{d_{2}}+e^{-i\int^{t}_{0}J_{2}(t^{\prime})dt^{\prime}\sin\phi_{2}(\ket{a_{2}}\bra{a_{2}}+\ket{b_{2}}\bra{b_{2}})}
\notag\\
&\times e^{-i\int^{t}_{0}J_{2}(t^{\prime})dt^{\prime}[-\cos\phi_{2}(\ket{a_{2}}\bra{b_{2}}+\ket{b_{2}}\bra{a_{2}})+
\sin\phi_{2}(\ket{a_{2}}\bra{a_{2}}-\ket{b_{2}}\bra{b_{2}})]}.
\end{align}
If the evolution period $T$ is taken to satisfy
\begin{align}
\int^{T}_{0}J_{2}(t)dt=\pi,
\end{align}
the evolution operator is reduced to
\begin{align}\label{eq8}
U_{2}(T)=&\ket{d_{1}}\bra{d_{1}}
+e^{-i(\pi+\pi\sin\phi_{2})}\ket{b_{1}}\bra{b_{1}}
\notag\\
&+\ket{d_{2}}\bra{d_{2}}
+e^{-i(\pi+\pi\sin\phi_{2})}\ket{b_{2}}\bra{b_{2}}
\notag\\
&+e^{-i(\pi+\pi\sin\phi_{2})}\ket{a_{1}}\bra{a_{1}}+e^{-i(\pi+\pi\sin\phi_{2})}\ket{a_{2}}\bra{a_{2}}.
\end{align}
From Eqs. (\ref{eq7}) and (\ref{eq8}), we can see that a quantum state initially residing in the computation space  $\mathcal{S}_{2}=\mathrm{Span}\{\ket{00}_{L},\ket{01}_{L},\ket{10}_{L},\ket{11}_{L}\}$ will evolve outside $\mathcal{S}_{2}$ during $t\in(0,T)$ and finally return back to $\mathcal{S}_{2}$ at $t=T$, i.e., the cyclic evolution condition (i) is satisfied. By using the commutation relation $[H_{2}(t),U_{2}(t)]=0$, we can verify that $\bra{\psi(t)}H_{2}(t)\ket{\psi(t)}
=\bra{\psi(0)}U^{\dagger}_{2}(t)H_{2}(t)U_{2}(t)\ket{\psi(0)}
=\bra{\psi(0)}H_{2}(t)\ket{\psi(0)}=0$, where $\ket{\psi(t)}$ is an evolution state such that $\ket{\psi(t)}=U_{2}(t)\ket{\psi(0)}$ with $\ket{\psi(0)}\in\mathcal{S}_{2}$. It means that the parallel transport condition (ii) is satisfied. Therefore, $U_{2}(T)$ is a holonomic transformation. When $U_{2}(T)$ acts on the computational space, it is equivalent to
\begin{align}
U_{2}=&\ket{d_{1}}\bra{d_{1}}
+e^{-i(\pi+\pi\sin\phi_{2})}\ket{b_{1}}\bra{b_{1}}
\notag\\
&+\ket{d_{2}}\bra{d_{2}}
+e^{-i(\pi+\pi\sin\phi_{2})}\ket{b_{2}}\bra{b_{2}},
\end{align}
which plays the role of a nonadiabatic holonomic gate. Similar to one-qubit gates, we can demonstrate that the two-qubit gate $U_{2}$ can be also protected by dynamical decoupling.

In the following, we demonstrate that $U_{2}$ is a nontrivial two-qubit gate.
We can see that $U_{2}$ can be rewritten as
\begin{align}
U_{2}=&\ket{0}_{LL}\bra{0}\otimes e^{-i\gamma_{2}(\sin\theta_{2}X+\cos\theta_{2}Z)/2}
\notag\\
&+\ket{1}_{LL}\bra{1}\otimes e^{-i\gamma_{2}(\sin\theta_{2}X-\cos\theta_{2}Z)/2}
\end{align}
with $\gamma_{2}=-(\pi+\pi\sin\phi_{2})$. Here, an unimportant global phase has been ignored.
Obviously, $U_{2}$ is a nontrivial two-qubit gate.
This two-qubit gate can realize a frequently used controlled phase gate when assisted by a one-qubit gate. Specifically, we first set $\theta_{2}=0$, and the two-qubit gate is reduced to $U_{2}=\ket{0}_{LL}\bra{0}\otimes\exp(-i\gamma_{2}Z/2)+\ket{1}_{LL}\bra{1}\otimes\exp(i\gamma_{2}Z/2)$. We then combine this gate and the one-qubit gate $\exp(i\gamma_{2}Z/2)$ acting on the second logical qubit, and the controlled phase gate $U_{C-P}=\ket{0}_{LL}\bra{0}+\ket{1}_{LL}\bra{1}\otimes\exp(i\gamma_{2}Z)$ can be realized.

\section{Conclusion}

In conclusion, we have put forward a protocol of nonadiabatic holonomic quantum computation protected by dynamical decoupling. A universal set of dynamical-decoupling-protected nonadiabatic holonomic gates is realized. Considering that geometric phases are only robust against control errors but cannot resist environment-induced decoherence, dynamical decoupling indeed provides an effective method to reduce the influence of the environment on nonadiabatic holonomic gates.
Our protocol can protect nonadiabatic holonomic gates against both collective decoherence and independent decoherence. Due to the combination of nonadiabatic holonomic quantum computation and dynamical decoupling, our protocol not only possesses the intrinsic robustness against control errors but also protects quantum gates against environment-induced decoherence.

\begin{acknowledgments}
This work was supported by the National Natural Science Foundation of China through Grants No. 11947221 and No. 11775129.
\end{acknowledgments}

\end{document}